\documentstyle[12pt,epsfig]{article}
\newcommand{\ccaption}[2]{
    \begin{center}
    \parbox{0.85\textwidth}{
      \caption[#1]{\small\it {#2}}}                   
    \end{center}    }            
\def    \be             {\begin{equation}}
\def    \ee             {\end{equation}}
\def    \ba             {\begin{eqnarray}}
\def    \ea             {\end{eqnarray}}
\def    \nn             {\nonumber}
\def    \=              {\;=\;}
\def    \frac           #1#2{{#1 \over #2}}

\def    \to             {\rightarrow }
\def    \et             {\mbox{$E_{\rm T}$}}

\def    \zprime         {\mbox{$V$}}
\def\gamzp{\mbox{$\Gamma_{V}$}}         
\def\mzp{\mbox{$M_{V}$}}      
\def\mzpsq{\mbox{$M_{V}^2$}}    
\def\as{\mbox{$\alpha_s$}}  
\def\ap{\mbox{$\alpha'$}}
\newcommand{\bea}{\begin{eqnarray}}
\newcommand{\eea}{\end{eqnarray}}

\begin{document}
\begin{titlepage}
\nopagebreak
{\flushright{
        \begin{minipage}{6cm}
        CERN-TH/96-20\hfill\\
        UGVA--DPT 1996/01--912\hfill\\
        hep-ph/9601324\hfill \\
        \end{minipage}        }

}
\vfill                        
\begin{center}

{\LARGE { \bf \sc   $R_b$, $R_c$ and Jet Distributions \\[0.2cm]
  at the Tevatron in a Model \\[0.2cm]
  with an Extra Vector Boson \footnote{Work partially supported by the Swiss
  National Foundation.}\\}}      
\vfill                                
{Guido ALTARELLI$^{\,a,b}$, Nicola DI BARTOLOMEO$^{\,c}$, \\
Ferruccio FERUGLIO$^{\,d}$, 
Raoul GATTO$^{\,c}$ and     \\   
Michelangelo L. MANGANO$^{\,a}$\footnote{On leave of absence from 
INFN, Pisa, Italy}   }
\vskip .3cm                               
{\it 
$^a$ CERN, Theory Division, \\
   1211 Geneva 23, Switzerland \\
$^b$ Dipartimento di Fisica, Universita' di Roma III, Italy \\
$^c$ D\'epartement de Physique Th\'eorique, Universit\'e de \\
          Gen\`eve, CH-1211 Geneva 4, Switzerland \\          
$^d$ Dipartimento di Fisica, Universita' di Padova, and \\
          INFN, Sezione di Padova, Italy \\}
\end{center}                                                      
\nopagebreak
\vfill                                
%\vskip 3cm                                               
\begin{abstract}
We show that the reported anomalies in $R_b$ and $R_c$ 
can be interpreted as the effect of a heavy vector boson \zprime\
universally coupled to $u$- and $d$-type quarks separately and nearly decoupled
from leptons. This extra vector boson could then also naturally explain the
apparent excess of the jet rate at large transverse momentum observed at CDF.
\end{abstract}                                                               
\vskip 1cm
CERN-TH/96-20\hfill \\
January 1996 \hfill
\vfill      
\end{titlepage}

\section{Introduction}
On the whole the electroweak (EW) precision tests performed at LEP, SLC and at
the Tevatron have impressively confirmed the formidable accuracy of the
Standard Model (SM) predictions. There are only a few 
hints of                         
possible deviations and our hopes of finding new physics signals are confined
to them. At LEP the observed values of $R_b$ and $R_c$ deviate from the SM
predictions by about 3.5$\sigma$ and 2.5$\sigma$, respectively \cite{eewg95}. 
At CDF an
excess of jets at large \et\ with respect to the QCD prediction has been 
reported~\cite{cdfjet}. 
None of these observations provides a very compelling evidence 
for new physics as yet, because of the limited statistics and of possible 
residual experimental systematics.
The $R_b$ value is relatively more established, in the sense that it was first
announced in 1993 and is insofar supported by the analyses of all four LEP
collaborations with several independent, in principle clean, tagging methods.
From a speculative point of view it is not implausible to have a deviation in
the third generation sector. Also, a moderate increase of $R_b$ with respect to
the SM (of roughly half of the present excess) would bring the value of
$\alpha_s(M_Z)$ measured from the $Z$ widths in even better agreement with lower
energy determinations. The $R_c$ evidence is much less believable both from the
experimental and the theoretical points of view. In absolute terms it is a 
large                                                              
deficit, that would overcompensate the $R_b$ excess. Thus these
results, if taken at face value, would demand a deviation from the SM in the
light-quark widths as well, in order to reestablish the observed value of
$\Gamma_h$, which is measured with great experimental accuracy and agrees with
the SM. After all, in this context charm is alike any other first or
second generation quark, while beauty could be special, being connected to
the heavy top. If one literally believes the data, then one must accept 
an accurate cancellation among the new physics contributions to
light and heavy quarks. But the perfect agreement of the leptonic widths 
with the SM, up to a fraction of MeV, clearly poses the problem of how
to naturally shift the light quark widths without affecting the leptonic ones as
well. Finally, the significance of the CDF result on jets 
entirely depends on the calculation of the QCD predictions at
large \et, which could to some extent be questioned. For example, it was
recently pointed out \cite{tung}\ that it is possible to 
slightly increase the large-$x$ gluon densities without deteriorating the
standard overall fits to low energy data, and thus partly explain a large
fraction of the high-\et\ jet discrepancy.

All these words of caution being said, in this note we consider the challenging
task of quantitatively explaining in an admittedly ad hoc but relatively simple
model all the three observed deviations discussed above.
We introduce a heavy vector neutral resonance
\zprime, singlet with respect to the standard gauge group $SU(3) \times SU(2)_L 
\times U(1)_Y$ and with a mass in the TeV range. We allow
 this new resonance $V$ to have a small mixing with the ordinary $Z$ gauge
 boson, and therefore to contribute to the $Z$ decays. 
We observe that while in the data $\delta(R_b+R_c)$ is large and negative,
$\delta(3R_b+2R_c)$ is only about $1\sigma$ away from zero. This suggests to
take universal couplings of the \zprime\ to the three generations of fermions
separately for up, down and charged leptons. Since the leptonic width is
in perfect agreement with the SM,
the leptonic couplings of \zprime\ must be much smaller than those needed
for the quarks to explain the deviations observed via $\delta R_b$
and via $\delta R_c$, and we shall take them as approximately vanishing (at a
less phenomenological level, one must be prepared to add new, presumably very
heavy, fermions to compensate the anomalies). Then the products of the amount
of mixing (which is severely constrained by the data) times the couplings of
the \zprime\ to up- and down-type quarks are fixed by imposing that
the observed values of $\delta R_b$ and of $\delta(3R_b+2R_c)$ be
approximately reproduced.
We have at our disposal five parameters to do that: the amount of mixing, 
\mzp\ (that for a
given mixing fixes $\epsilon_1=\delta\rho$), the left-handed coupling to the
$(u,d)_{\rm L}$ doublets, and the two right-handed couplings to the $u_{\rm R}$
and $d_{\rm R}$ singlets. So the game would be trivial, were it 
not for the fact that
couplings to quarks as large as those required by $\delta R_b$ and $\delta R_c$
would tend to produce too large effects in the distributions of 
large-\et\ jets measured at the Tevatron. We can then adjust $\mzp\ge 1$~TeV 
and the left and right couplings in such a way as to obtain a reasonable fit to
both  LEP and CDF anomalies, without violating, to our knowledge, any known
experimental constraint. The details are given in what follows.                     

\section{Effects on LEP and SLC observables}

The tree level neutral current
interaction can be written in terms of the unmixed 
interaction states $Z_0$ and
$V_0$, coupled respectively to the ordinary standard model neutral
current
$(J_{3L} - \sin^2\theta_W J_{em})$ and to an additional current $J_N$.
The vector and axial couplings of the gauge bosons $Z_0$ and $V_0$  
are defined by:
\bea
{\cal L}_{NC} & = & \frac{g}{2 \cos\theta_W}\sum_{i} \left[ Z_0^{\mu}(
v_S^i {\bar \psi}^i \gamma_\mu \psi^{i} + a_S^{i}
{\bar \psi}^i \gamma_\mu \gamma_5 \psi^{i} ) \right. \nonumber \\
& + & \left. V_0^{\mu}(
v_N^i {\bar \psi}^i \gamma_\mu \psi^{i} + a_N^{i}
{\bar \psi}^i \gamma_\mu \gamma_5 \psi^{i} ) \right]
\label{2}
\eea

The $Z_0$ couplings are the standard ones
\be
v_S^{i} = T^i_{3L} - 2 \sin^2\theta_W Q^i\; , ~~~ a_S^i = - T^i_{3L}
\label{3}
\ee
where $T^i_{3L}$ is the third component of the weak isospin of the
fermion
$i$, and $Q^i$ its electric charge.

We assume that the new gauge boson $V_0$ couples only 
to the quarks and has zero (or negligible) couplings
to the leptons. We also assume family-independent couplings. 
The new interactions can be then  be expressed in terms of three parameters
$x$, $y_u$ and $y_d$:                          
\bea
v_N^{u} = x + y_u ~~ & , & ~~ a_N^{u} = -x + y_u \nonumber \\
v_N^{d} = x + y_d ~~ & , & ~~ a_N^{d} = -x + y_d,
\label{4}                                        
\eea
where the superscripts $u$ and $d$  refer  to up-type and down-type
quarks.

In presence of a mixing, the mass eigenstates $Z$ and $V$ 
are given by a rotation of the 
unmixed states $Z_0$ and $V_0$:
\bea
Z & = & \cos \xi Z_0 + \sin \xi V_0 \nonumber \\
V & = & -\sin \xi Z_0 + \cos \xi V_0
\label{5}
\eea

Due to the mixing, the $\rho$ parameter, defined by
\be
\rho = \frac{M_W^2}{M_Z^2 \cos^2\theta_W}
\label{6}
\ee
receives a tree level contribution $\Delta \rho_M$, which in term of
the $V$ mass and the mixing angle $\xi$ is given by:
\be
\Delta\rho_M = \left[ \left(\frac{M_{V}}{M_Z} \right)^2 -1 \right]
\sin^2\xi \simeq \left( \frac{\mzp}{M_Z} \right)^2 \, \xi^2
\label{7}                                               
\ee

At LEPI the observables get corrections from the presence
of $V$ through the mixing with the ordinary $Z$ and
through the
shift in the $\rho$ parameter. Contributions from direct $V$
exchange are negligible at the $Z$ pole, but will be taken into
account                                              
later on in our study of the Tevatron jet observables.
                                          
The deviation  of a LEPI observable, linearized in $\Delta\rho_M$ and $\xi$,
can therefore be expressed as:
\be                           
\frac{\delta {\cal O}}{{\cal O}} = A_{\cal O} \Delta\rho_M + B_{\cal O}
\xi.
\label{10}
\ee

The coefficients $A_{\cal O}$ are universal and
depend only on the SM parameters and
couplings, while $B_{\cal O}$ also depend on the $V_0$ couplings
$v^i_N$ and                               
$a^i_N$~\cite{alt90}. In Table~I we give the numerical values of                  
$A_{\cal O}$ and the expressions for $B_{\cal O}$ 
for the observables of interest, as functions of the parameters $x$, $y_u$                            
and $y_d$ introduced in eq.~(\ref{4}). 
In Table~I we also present the experimental data 
used in the present analysis \cite{eewg95}, together with the
Standard Model predictions \cite{repprec} for $m_{top} = 175$~GeV,
$m_H = 300$~GeV and  $\alpha_s(M_Z) = 0.125$.                  
They include the one-loop electroweak radiative corrections.
The $Z$ mass was fixed at the
experimental value $M_Z = 91.1887$~GeV.
\\[0.1cm]
% TABELLA  A   B                      
{\begin{center}
\footnotesize
\begin{tabular}{|c||r|c|c|r|r|}   
\hline                        
& & & & & \\
Quantity & $A$ & $B$ & Exp. values~\cite{eewg95} 
& SM values & Pull of the fit \\
& & & & & \\                                                  
\hline\hline
& & & & & \\
${\Gamma_Z}$ & 1.36 & $ -0.92 x -0.49 y_u + 0.37 y_d $ 
& $2496.3 \pm 3.2$  & 2497.4 & 1.72 
\\                  
$R_l = \Gamma_h/\Gamma_l$ & 0.34 & $ -1.31 x -0.70 y_u + 0.52 y_d $ 
& $20.788 \pm 0.032$ & 20.782 & $-0.35$ 
\\        
$\sigma_h  $ & $-0.030$ & $ 0.52 x +0.28 y_u -0.21 y_d $ 
& $41.488 \pm 0.078$ & 41.451 & $-0.32$                    
\\
$R_b = \Gamma_b/\Gamma_h$ & $-0.094$ & $ -3.16 x + 0.70 y_u + 0.29 y_d $               
& $0.2219 \pm 0.0017$ & 0.21569  & $-1.41$
\\                                                          
$R_c = \Gamma_c/\Gamma_h$ & 0.12 & $ 6.20 x - 1.43 y_u - 0.59 y_d $ 
& $0.1543 \pm 0.0074$ & 0.17238 & 1.62 
\\
$M_W/M_Z$ & 0.71 & 0 
& $0.8802 \pm 0.0018$ & 0.8808 & 0.94 
\\                     
${\cal A}_l$ & 18.50 & 0 
& $-0.15066 \pm 0.00276$ & $-0.14334$& 0.98
\\                 
${\cal A}_b$ & 0.23 & $ -0.31 x - 1.72 y_d $ 
& $-0.841 \pm 0.053$ & $-0.9342$& $-1.79$ \\
${\cal A}_c$ & 1.70 & $ 2.37 x + 5.36 y_u $ 
& $-0.606 \pm 0.090$ & $-0.6662$ & $-1.00$ \\
$A_{FB}^{b}$ & 18.15 & $ -0.31 x - 1.72 y_d $ 
& $0.0999 \pm 0.0031$ & 0.10042 & 1.20\\
$A_{FB}^{c}$ & 19.63 & $ 2.37 x + 5.36 y_u $ 
& $0.0725 \pm 0.0058$ & 0.07161 & 0.76  \\
& & & & & \\
\hline                        
\end{tabular}                             
\end{center}}
\vspace{3mm} 
{\bf Table~I} : Coefficients $A$ and $B$, defined in eq. (\ref{10}), for
various electroweak observables, together with their experimental values and SM
theoretical predictions for $m_{top} = 175 \; GeV$,
$m_H = 300 \; GeV$ and  $\alpha_s(M_Z) = 0.125$. 
The corresponding $\chi^2$ is equal to 26.73.
In the last column we report the pull values                    
((fit-exp)/$\sigma$) for the final \zprime\ fit with $x=-1$, $y_u=2.2$, $y_d=0$
and $\xi = 3.8 \cdot 10^{-3}$. The $\chi^2$ in this case equals 14.72.
\vspace{0.5cm}                                                      

The deviations in Table I are computed from the tree level
formulas for the partial widths
\be
\Gamma (Z \to {\bar f}f )= \frac{G_F M_Z^3}{6 \pi \sqrt{2}} \rho N_c
\left[
(v_{eff}^f )^2 + (a_{eff}^f )^2 \right],
\label{11}
\ee
and for the asymmetries
\be
{\cal A}_f = \frac{2 a_{eff}^{f} v_{eff}^{f}}{
(v_{eff}^f )^2 + (a_{eff}^f )^2}.
\label{12}
\ee
The forward-backward asymmetries are given by:
\be
A_{FB}^{f} = \frac{3}{4} {\cal A}_e {\cal A}_f
\label{13}
\ee
In eq.~(\ref{11}) $N_c = 3$ for quarks and $N_c = 1$ for leptons, and
in eq.~(\ref{11}) and (\ref{12}) the effective vector and axial-vector
coupling
$v_{eff}^{f}$ and $a_{eff}^{f}$  are superpositions of the corresponding
$Z_0$ and $V_0$ couplings:
\bea
v_{eff}^{f} = \cos\xi \, v_S^f + \sin\xi \, v_N^f \nonumber \\
a_{eff}^{f} = \cos\xi \, a_S^f + \sin\xi \, a_N^f
\label{14}                                  
\eea

In computing the deviations due to the new vector resonance $V$,
it is sufficient to consider
the tree level expressions for the observables, because the corrections
are
proportional to $\Delta\rho_M$ or $\xi$, that are both constrained to be
quite small (of the order $10^{-3}$) by the current electroweak data.
                                                    
We keep fixed the input
parameters $\alpha$, $G_F$, $M_Z$, and take into
account the modification of            
the effective Weinberg angle given in eq.~\ref{6}
because of the shift in the $\rho$ parameter. One finds \cite{alt90}:
\be                                                                  
\delta (\sin^2\theta_W ) = - \frac{\sin^2\theta_W\cos^2\theta_W}{
\cos^2 2\theta_W}\Delta\rho_M
\label{16}                   
\ee

The loop effects due to the heavy gauge boson $V$ are quite small and
we will
neglect them.

\section{Fit to the LEP and SLC data}

In this Section we constrain the free parameters of our extended gauge
model
by performing  a fit of the eleven independent observables of Table I.
The parameter space of the model includes the couplings $x$, $y_u$, $y_d$ 
and  the two  parameters  $\xi$ and $M_{V}$.  
 $\Delta\rho_M$ is related to the previous parameters by eq.~(\ref{7}).

We have minimized the $\chi^2$ function keeping $M_{V}$ fixed at
different values: it turns out that
 the best fit central value for
 $\Delta\rho_M$ stays almost
fixed, by varying $M_{V}$,  at the value
\be                                     
\Delta\rho_M \simeq 0.0011
\label{18}
\ee
This implies, from eqs.~(\ref{7}), that 
the mixing angle $\xi$ decreases with $M_{V}$:
\be
\xi \simeq \sqrt{0.0011} \frac{M_{Z}}{M_{V}}
\label{19}
\ee

The parameters $x$, $y_u$, and $y_d$ are multiplied by the mixing angle
$\xi$
in the expression~(\ref{10}) for the deviations: that means, from
eq.~(\ref{19}),       
that their best fit values will scale with $M_{V}$ as $1/\xi$, i.e.
\be
x,~y_u,~y_d \sim \frac{M_{V}}{M_Z}
\label{20}
\ee

The fit, for a choice $M_{V} = 1000$~GeV, leaving the four
parameters                                
$\xi$, $x$, $y_u$, and $y_d$ as free, gives:
\bea
\xi = (3.0^{+ 0.9}_{- 1.2}) \cdot 10^{-3} & , & ~ x = -1.4^{+1.1}_{-1.9} \nonumber \\
y_u = 5.3^{+4.1}_{-2.1} & , & ~ y_d = 2.9^{+ 4.9}_{- 4.6}
\label{21}
\eea

We have quoted the standard errors, corresponding to $\chi^2 =
\chi^2_{min} +1 $.
The fit is weakly sensitive to the parameter $y_d$, as the large error
indicates. We take advantage of this by constraining the fit with $y_d=0$.
The other parameters of the fit with $y_d=0$ turn out to be:
\be                                          
\xi = (2.8^{+0.9}_{-1.4}) \cdot 10^{-3}~ ,
 ~ x = -2.1^{+0.8}_{- 2.4} ~ , ~ y_u = 4.5^{+4.4}_{-1.5}
\label{22}
\ee
where we have fixed, as before, $M_{V} = 1000$~GeV. 
Here $\chi^2 = 11.41$. For other values of
$M_{V}$, the scaling formulas eq.~(\ref{19}) and (\ref{20}) are an
excellent approximation. The central values correspond to quite large couplings
that would be incompatible with the CDF data, as shown in the next Section.
                                                                           
The parameters in eq.~(\ref{22}) are strongly correlated. 
The correlation between the parameters $x$ and $y_u$ is easily understood
once noticed, from Table I, that the ratio of the coefficients multiplying
$x$ and $y_u$ in the formulas for the deviations is the same, 1.87,
 in the observables 
$\Gamma_Z$, $R_l$ and $\sigma_h$. The relative high precision data are in
excellent agreement with the SM predictions, and this induces a strong
anticorrelation between the two parameters.
  
\begin{figure}                                                           
\centerline{
\epsfig{figure=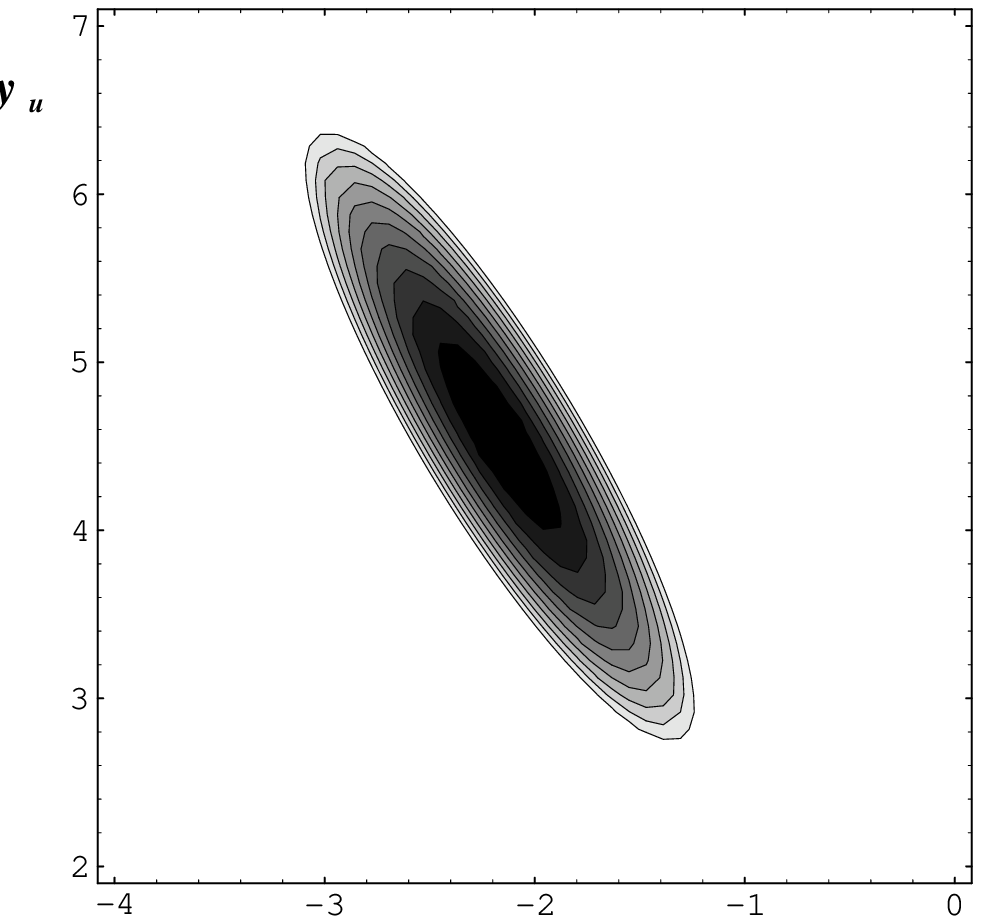,width=0.4\textwidth,angle=0} \hfil
\epsfig{figure=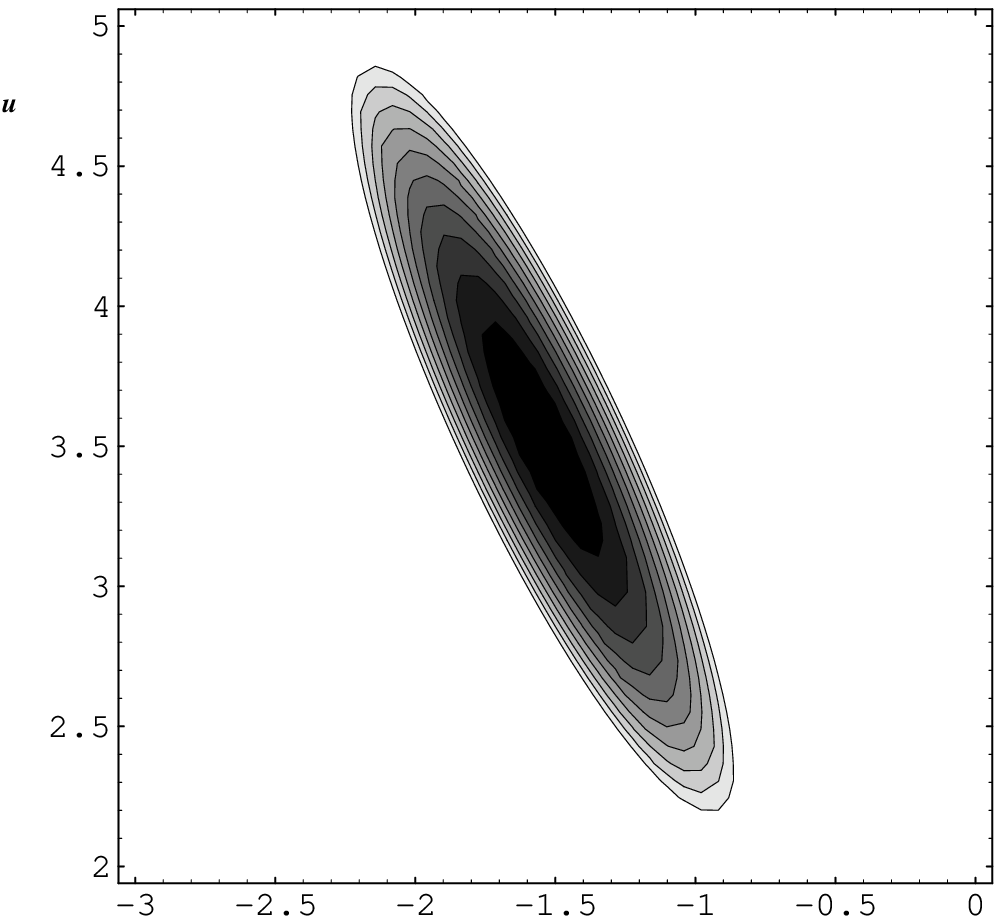,width=0.4\textwidth,angle=0}} 
\ccaption{}{\label{ffit}                           
$70 \%$ confidence level ellipsis in the plane $y_u$ 
          versus $x$, for $M_{V} = 1000$ GeV, $y_d = 0$ and 
$\xi=2.8 \cdot 10^{-3}$ (left
figure) or $\xi=3.8 \cdot 10^{-3}$ (right figure).}                                
\end{figure}

In fig.~\ref{ffit}a we
plot                                                             
the $70\%$ confidence level ellipsis in the plane $y_u$ versus $x$,
keeping
$\xi$ fixed at the best-fit value $2.8 \cdot 10^{-3}$. From the figure, 
one can see that
at this
confidence level the points closest to the origin are at the position
$x \simeq -1.3$, $y_u \simeq 2.8$.

In fig.~\ref{ffit}b we present the analogous ellipsis for the higher value
$\xi = 3.8 \cdot 10^{-3}$.
Increasing $\xi$, the elliptical region moves toward the origin, because
the higher
mixing angle  forces the parameters $x$, $y_u$ to smaller values.
For the value of $\xi = 3.8 \cdot 10^{-3}$ the closest 
points to the origin are located
at $x \simeq -1.0$, $y_u \simeq 2.2$.
Moving away from the best-fit value of $\xi$, the $\chi^2$ value
increases:
for $\xi = 3.8 \cdot 10^{-3}$, $ x = -1.0$, and $y_u = 2.2$, one obtains 
$\chi^2 = \chi^2_{min}
+ 3.3$, still in the $70 \%$ confidence level region of the three
parameters
fit of eq.~(\ref{22}).
 
In the last column of Table~I we quote the pull values, given by
(fit - exp)/$\sigma$, for $\xi = 3.8 \cdot 10^{-3}$, $x = -1$, $y_u = 2.2$
and $y_d = 0$: the discrepancies in $R_b$ and $R_c$ are reduced.
We stress again
that these are not the best fit values for the parameters, but they 
 lead to an effect on jet observables which is
quite compatible with the CDF observations, as we shall now discuss.
 
\section{Comparison with the Tevatron jet distributions}

A vector resonance \zprime\ with such large
 couplings as obtained from the fits of the previous
Section is liable to produce visible effects in hadronic collisions. There it
can be directly produced via the Drell-Yan mechanism if the mass is not too
large, or can lead to effective interactions between quarks via virtual
exchange. The net result is a growth of the inclusive \et\ distribution of jets
at large \et, relative to the standard QCD expectations. Using the \zprime
couplings defined in eq.~(\ref{4}), it is easy to evaluate the following 
quark-quark scattering amplitudes (amplitudes for crossed channels can be
easily obtained from these ones):
\ba                                        
\frac{1}{(4\pi)^2}&&\overline{\sum} \vert A (qq \to qq) \vert^2 \=
\frac{1}{(4\pi)^2}\overline{\sum} \vert A_{QCD} (qq \to qq) \vert^2 +
\nn \\                                    
&& \frac{32}{9}\as\ap s^2 {\cal{R }}\bigl\{\frac{1}{t[(u-\mzpsq)+i\mzp\gamzp]}  
       +                                   
   \frac{1}{u[(t-\mzpsq)+i\mzp\gamzp]} \bigr\} (x^2+y_q^2) +
\nn \\                                                                     
&& 16\ap^2 \Bigl\{ s^2 (x^4+y_q^4)\bigl[\frac{1}{(t-\mzpsq)^2+\mzpsq\gamzp^2} +
                       \frac{1}{(u-\mzpsq)^2+\mzpsq\gamzp^2} +
\nn \\                                                                     
&&
  \frac{2}{3}{\cal{R }}
   \frac{1}{(t-\mzpsq)+i\mzp\gamzp}\frac{1}{(u-\mzpsq)-i\mzp\gamzp} \bigr]
  +                                                                
\nn \\                                                                     
&&
      2x^2y_q^2 \bigl[\frac{u^2}{(t-\mzpsq)^2+\mzpsq\gamzp^2}+             
              \frac{t^2}{(u-\mzpsq)^2+\mzpsq\gamzp^2} \bigr] \Bigr\}
\\[0.2cm]                                                                  
\frac{1}{(4\pi)^2}&&\overline{\sum} \vert A (qq'\to qq') \vert^2 \=
\frac{1}{(4\pi)^2}\overline{\sum} \vert A_{QCD} (qq' \to qq') \vert^2 +
\nn \\                                                                     
&&
\frac{16\ap^2}{(t-\mzpsq)^2+\mzpsq\gamzp^2}                 
    \bigl [ s^2\,(x^4+y_q^2y_{q'}^2)+u^2 x^2\,(y_q^2+y_{q'}^2)\bigr]            
\ea                                                                
where $A_{QCD}$ is the standard QCD amplitude, ${\cal R}$ denotes the real
part, $\ap=g^2/(16\pi\cos^2\theta_W)\sim 0.010$
and $\gamzp$ is the \zprime\ total                                             
decay width given by:
\be
\gamzp = 2N_g\ap\mzp(2x^2+y_u^2+y_d^2)
\ee
Taking the $\mzp\to\infty$ limit one recovers\footnote{Up to some misprint
contained in the standard literature.} the standard results obtained in
presence of an effective 4-quark coupling \cite{peskin} .          
                                                                        
\begin{figure}                                                           
\centerline{
\epsfig{figure=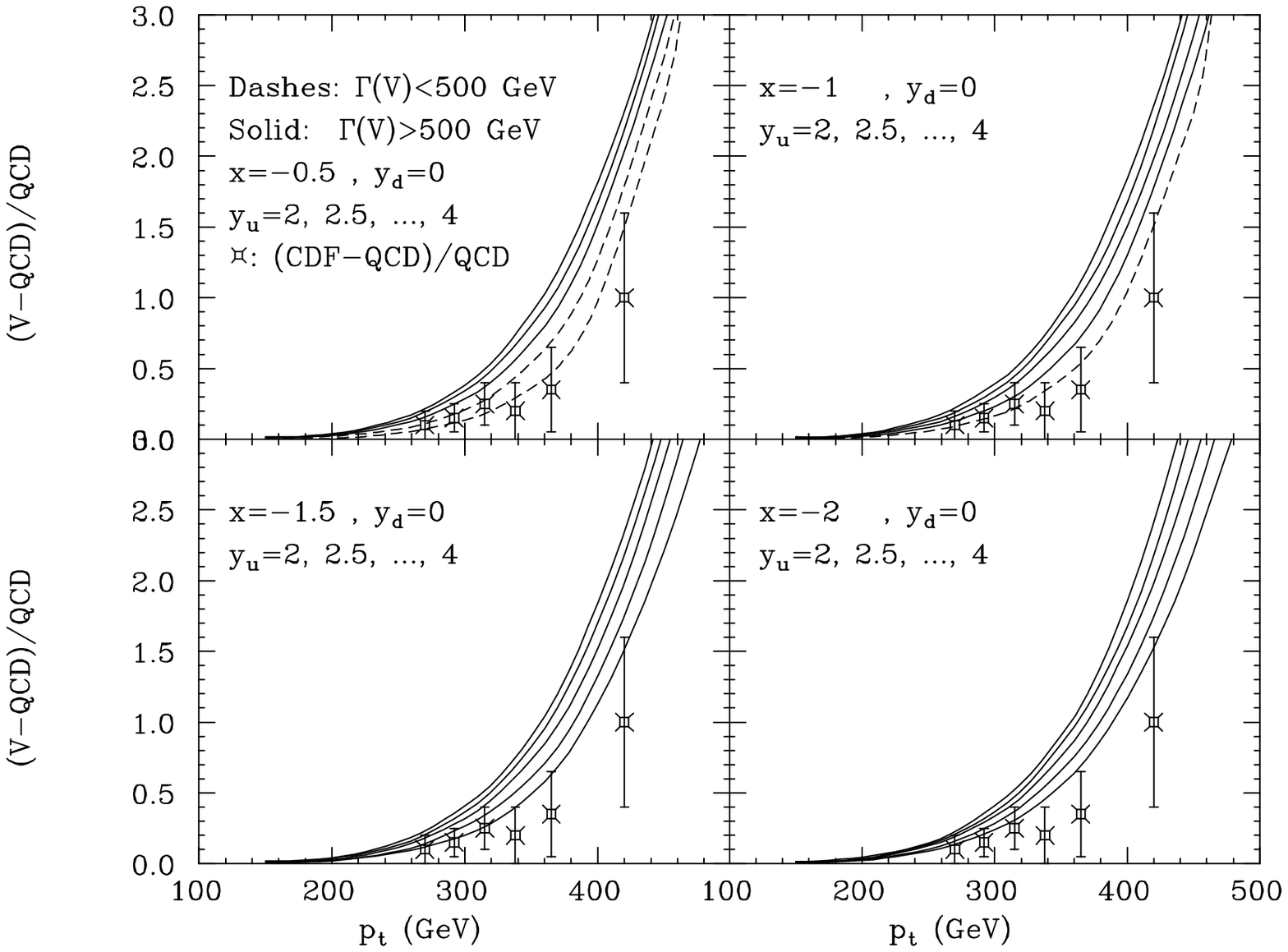,width=0.8\textwidth,angle=0}} 
\ccaption{}{\label{figcdf1}                   
The effect of \zprime\ exchange on the inclusive \et\ distribution of jets at
the Tevatron. The different curves correspond to increasing values of $y_u$,
from 2 to 4. The four displays correspond to $x=-0.5$, --1, --1.5 and --2. The
quantity $(CDF-QCD)/QCD$~\cite{cdfjet}                                  
is shown by the points. We used dashed or continuous lines to indicate whether
the \zprime\ width is smaller or larger than 500~GeV.}
\end{figure}

\begin{figure}                                                           
\centerline{
\epsfig{figure=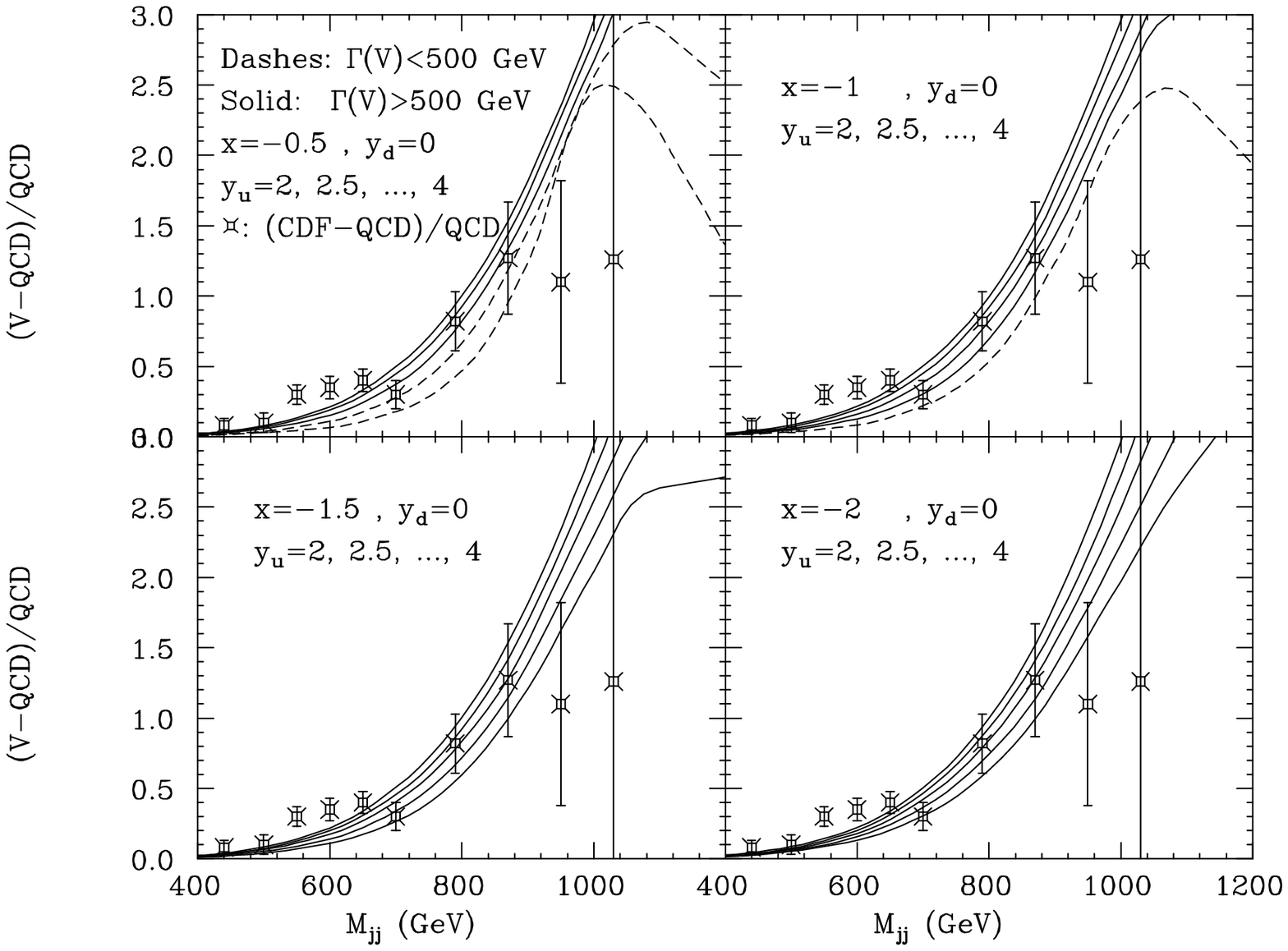,width=0.8\textwidth,angle=0}}
\ccaption{}{\label{figcdf2}                   
The effect of \zprime\ exchange on the invariant mass distribution of di-jet
events 
at the Tevatron. The different curves correspond to increasing values of $y_u$
from 2 to 4. The four displays correspond to $x=-0.5$, --1, --1.5 and --2. The
quantity $(CDF-QCD)/QCD$~\cite{cdfmass}
is shown by the points.}
\end{figure}

Fig.~\ref{figcdf1}\ shows the deviations induced by the couplings to the
\zprime\ on the jet inclusive \et\ distribution at the Tevatron. The quantity:
\be
     \bigl[\frac{d \sigma^{(QCD+V)}/d\et}{d \sigma^{QCD}/d\et}\bigr]_{\eta=0}
      -1
\ee                                             
is plotted as a function of jet \et\ for different values of $x$ and $y_u$,
chosen in the range favoured by the EW fits.  This is compared to the CDF 
data~\cite{cdfjet}, represented in the figure as:
\be                                              
     \frac{d \sigma^{CDF}/d\et}{d \sigma^{QCD}/d\et}-1
\ee                                             
The calculation of the \zprime\ contribution incorporates 
the full                                     
set of QCD processes, including reactions initiated by $gg$ and $gq$. Only LO
diagrams are considered, as no NLO calculation for the \zprime-exchange 
contribution is available. The calculation was performed using the MRSA set of
parton densities, and a renormalization scale $\mu=\et$. We verified that
the quantity displayed in fig.~\ref{figcdf1}
is very stable under changes of these parameters. We also expect that NLO
corrections should not affect significantly our results.

As the figure shows, the extreme choice
$x=-1$, $y_u=2.5$ allowed by the EW fits is fully consistent with the CDF
data. 
A similar conclusion can be reached by examining the di-jet mass
distribution, shown in fig.~\ref{figcdf2}. Notice that the peak structure
disappears for too large couplings, as the convolution of the large width and
the falling parton luminosities smears away the resonance.
                                          
\section{Conclusions}
Deviations from the SM in $R_b$, $R_c$, and in CDF jets have been reported.
They do not yet constitute compelling evidences for new physics.
Nevertheless one may want to take them at their face values and look for
some new effect to explain them. We introduce, as a simplest object,
a new heavy singlet vector boson, with some mixing to the $Z$ and direct
couplings to quarks, the same for all up and the same for all down quarks, we
perform the overall fit to LEP data, and see whether we can also explain CDF 
jets. This is possible , within the errors, with a vector boson of
mass larger or of the order of 1~TeV, weakly mixed to the $Z$, but rather 
strongly coupled to the quarks. We do not attempt at this stage 
any deeper theoretical construction.

After completing this work we received a paper where similar ideas are
discussed \cite{chiappetta}.

\section{ADDENDUM: Low energy neutral-current data}

The data analyzed in the main body of this work do not include low-energy 
neutral current experiments. The present Addendum is devoted to size the impact
of deep inelastic neutrino scattering on the allowed region in the parameter 
space.

The relevant information is contained in table II, where,
with the same notations used above, we list
experimental data, SM expectations and deviations
for the four parameters $g_{L,R}^2$ and $\theta_{L,R}$
characterizing $\nu$-hadron scattering \cite{pdg}.
%\\[0.1cm]
{\begin{center}
\footnotesize
\begin{tabular}{|c||c|c|c|c|c|}   
\hline                        
& & & & &   \\
Quantity & A & B & Exp. values~\cite{pdg} 
& SM values & Pull of the fit \\
& & & & &  \\                                                  
\hline\hline
& & &  & & \\
$g_L^2$ & 2.71 & -0.45 $x$ & $0.3017\pm0.0033$ & 0.303  & 0.76\\
$g_R^2$ & -0.60 & $-9.33 y_u + 4.67 y_d $ 
&$0.0326\pm0.0033$ & 0.030  & -1.63\\
$\theta_L$ & -0.07 & $-1.04 x $ & $2.50\pm0.035$ & 2.46  & -0.79\\
$\theta_R$ & 0.0 & $ 0.50 y_u + 1.00 y_d$ & 
$4.58^{+0.46}_{-0.28}$ &5.18  & 1.64 \\
\hline                        
\end{tabular}                             
\end{center}}
\vspace{3mm} 
{\bf Table~II} : Coefficients $A$ and $B$, defined as in eq. (2.7),
for low-energy neutral current observables, 
together with their experimental values and SM
theoretical predictions for $m_{top} = 175 \; GeV$,
$m_H = 300 \; GeV$.
% and  $\alpha_s(M_Z) = 0.125$. 
%The corresponding $\chi^2$ is equal to 26.73.
In the last column we report the pull values                    
((fit-exp)/$\sigma$) for $x=-2.1$, $y_u=4.3$, $y_d=0$
and $\xi = 2.3 \cdot 10^{-3}$. 
\vspace{0.5cm}                                                      

Including also the four low energy observables in the fit, fixing as 
before $M_V = 1000~GeV$ and $y_d= 0$ (the fit does not improve
significantly releasing this parameter) and leaving
the three parameters $\xi$, $x$ and $y_u$ free to vary,
one obtains:
\be
\xi = (2.3^{+1.1}_{-1.5}) \cdot 10^{-3}~  ,  ~ x = -2.1^{+1.0}_{-3.7}~ ,
~ y_u = 4.3 \pm 2.9
\label{a8}
\ee
The $\chi^2$ of the fit is 20.2, while the SM, for the values listed in
table II, gives $\chi^2 = 30.7$.
We recall that, by omitting the low-energy data, we obtained:
\be                                          
\xi = (2.8^{+0.9}_{-1.4}) \cdot 10^{-3}~ ,
 ~ x = -2.1^{+0.8}_{- 2.4} ~ , ~ y_u = 4.5^{+4.4}_{-1.5}
\label{a22}
\ee
Comparing (\ref{a8}) with (\ref{a22}), 
one notices that the low-energy data do not 
affect the results of the fit in any significant way:
 central values and errors are
essentially determined by the LEP data alone.

As we have discussed in the main body of the work, 
the central values in (\ref{a8}) or (\ref{a22}) 
give a too strong enhancement in the inclusive jet cross section at large
$E_T$, incompatible with the CDF data. 
In the low $\chi^2$ region, the values 
$\xi \simeq 3.8 \cdot 10^{-3}$, $x \simeq
-1.0$ and $y_u \simeq 2.2$ previously retained 
remain a good compromise also when including in the
fit the set of low energy data, which as we have shown  do not 
practically influence our analysis.

We have also included, in a following step, the weak charge $Q_W$ of Cesium
\cite{cesium} measured in atomic parity violation experiments: the result
is a small ($\sim 10 \%$) decrease 
of the central values of the parameters $x$ and $y_u$ in (\ref{22}).

In conclusion the situation remains practically unchanged after inclusion
in the fit of low energy data and we hope that future high energy data
will clarify the problem.

\par
\vspace*{1cm}
\noindent
{\bf Acknowledgements }\\
We thank Alain Blondel for the suggestion to include the neutrino deep 
inelastic data in our analysis, Paul Langacker for useful comments and 
Kevin McFarland for stimulating criticisms.

\end{document}